\documentclass[prd,twocolumn,aps,showpacs,nofootinbib,nobibnotes,superscriptaddress,preprintnumbers]{revtex4}
\usepackage{epsfig}
\usepackage{graphics}
\usepackage{bm}
\usepackage{color}
\usepackage{dcolumn}   
\usepackage{bm}     
\usepackage{bbm}       
\usepackage{amssymb}  
\usepackage{amsmath}
\usepackage{latexsym}
\usepackage{float}
\usepackage{ifthen}
\usepackage{caption,subfig}
\usepackage{enumerate}
\usepackage{url}
\usepackage{caption,subfig}
\usepackage{amsopn}

\usepackage{amsfonts}
\usepackage{multirow}
\usepackage{array}
\usepackage{booktabs}
\usepackage{rotating}

\usepackage{ulem}
\newcommand{\ud}{\mathrm{d}} 

\def\clap#1{\hbox to 0pt{\hss#1\hss}}

\def\({\left(}
\def\){\right)}
\def\[{\left[}
\def\]{\right]}
\def\bea{\begin{eqnarray}}
\def\eea{\end{eqnarray}}
\def\be{\begin{equation}}
\def\ee{\end{equation}}
\def\ba{\begin{eqnarray}}
\def\ea{\end{eqnarray}}
\def\beq{\begin{eqnarray}}
\def\eeq{\end{eqnarray}}

\newcommand{\cs}{c_s}

\def\cs{c_{\rm s}}

\def\be{\begin{equation}}
\def\ee{\end{equation}}
\def\ba{\begin{eqnarray}}
\def\ea{\end{eqnarray}}
\def\beq{\begin{eqnarray}}
\def\eeq{\end{eqnarray}}

\def\L*{{\cal L}_*}
\def\L{\mathcal{L}}
\def\({\left(}
\def\){\right)}

\def\<{\langle}
\def\>{\rangle}

\def\cs2{c_{s}^{2}}

\def\be{\begin{equation}}
\def\ee{\end{equation}}
\def\ba{\begin{eqnarray}}
\def\ea{\end{eqnarray}}
\def\beq{\begin{eqnarray}}
\def\eeq{\end{eqnarray}}

\def\L*{{\cal L}_*}
\def\L{\mathcal{L}}
\def\({\left(}
\def\){\right)}

\def\<{\langle}
\def\>{\rangle}

\begin{document}
\hspace{5.2in} \mbox{NORDITA-2015-52}\\\vspace{1.53cm} % Preprint number

\title{More on effective composite metrics}

\date{\today,~ $ $}

\author{Lavinia Heisenberg} \email{laviniah@kth.se}
\affiliation{Nordita, KTH Royal Institute of Technology and Stockholm
  University, \\Roslagstullsbacken 23, 10691 Stockholm, Sweden}
\affiliation{Department of Physics \& The Oskar Klein Centre, AlbaNova
  University Centre, 10691 Stockholm, Sweden}

\date{\today}

\begin{abstract}
  In this work we study different classes of effective composite metrics
  proposed in the context of one-loop quantum corrections in bimetric gravity. 
  For this purpose we consider contributions of the matter loops in form of cosmological constants 
  and potential terms yielding two types of effective composite metrics.
  This guarantees a nice behaviour at the quantum level. However, the
  theoretical consistency at the classical level needs to be ensured additionally.
  It turns out that among all these possible couplings only one unique effective
  metric survives this criteria
  at the classical level.

\end{abstract}

%PACS NEEDED

\maketitle

%-----------------------------------------------------------------
\section{Introduction}

The theoretical and observational challenges in modern cosmology,
like the cosmological constant problem, the recent acceleration
of the universe and dark matter have motivated many physicists 
to put in question the validity of General Relativity (GR) on cosmological scales. 
Many theories came alive, like scalar-tensor~\cite{Horndeski:1974wa, Nicolis:2008in,Deffayet:2009wt,
Deffayet:2009mn, deRham:2011by, Heisenberg:2014kea, Heisenberg:2014raa}, 
vector-tensor~\cite{Horndeski:1976gi,EPU10,PhysRevD.80.023004,
BeltranJimenez:2013fca,Jimenez:2013qsa,Heisenberg:2014rta,
Tasinato:2014eka} or tensor-tensor theories (in this context important
classes being massive gravity~\cite{deRham10, dRGT10}, bigravity~\cite{Hassan12a,
  Hassan12b} and multigravity~\cite{Hinter12} theories...etc).

Among all these possible modifications of GR, in this work we shall go along
the lines of bimetric theories. Since the past decade there has been a tremendous 
amount of effort to build a consistent, ghost-free, covariant and non linear
theory for massive gravity~\cite{deRham10, dRGT10}. It cousins like the bimetric ~\cite{Hassan12a, Hassan12b}
and multi-metric ~\cite{Hinter12} extensions share the same properties. Since then there
has been many interesting works on the theoretical and cosmological
implications of these theories, like the cosmological solutions~\cite{deRham:2010tw,PhysRevD.84.124046,PhysRevLett.109.171101}
(see also the references in~\cite{deRham:2014zqa}), consistent couplings to matter \cite{deRham:2014naa, deRham:2014fha, Heisenberg:2014rka, Hinterbichler:2015yaa, deRham:2015cha, Huang:2015yga}, 
possible new kinetic interactions \cite{Hinterbichler:2013eza,deRham:2013tfa, deRham:2015rxa},
applications to dark matter \cite{Blanchet:2015sra}...etc.

A ver nice property of massive gravity is its technical naturalness. Phenomenology requires
a small mass of the graviton. This smallness of the graviton mass is as badly tuned as for 
the cosmological constant. However, the explicit computations of the quantum corrections 
showed that the theory remains natural at the quantum level, meaning that the quantum 
corrections themselves are small \cite{deRham:2012ew, deRham:2013qqa}. 
In this context a natural question arose as whether
the matter fields can be coupled to both metrics of massive gravity simultaneously without
detuning the potential interactions at the quantum level. Upon this requirement an effective
composite metric built out of the two metrics was proposed in \cite{deRham:2014naa}. In a
follow-up work other types of possible effective metrics were considered \cite{Heisenberg:2014rka}. 
They constituted two classes of effective metric. The first class was constructed in a way that guaranteed the
quantum contributions to be in form of cosmological constants for the two metrics, which we
will denote here as $g_{\text{eff}}$. The second class of effective metrics were similar to the
one proposed in \cite{deRham:2014naa} in the sense that they give rise to quantum corrections
in the form of the allowed ghost-free potential interactions, represented by $\tilde{g}_{\text{eff}}$.
In the present work we will investigate whether these two types of effective metrics can constitute
consistent couplings at the classical level. For this purpose, we will first compute the Hamiltonian
of the mini-superspace and impose its linearity in the lapses. This is the first trivial test that these
couplings have to pass. As next, we will study the requirements for the ghost freedom in the
decoupling limit. Both analysis indicate that the effective metric proposed in \cite{deRham:2014naa}
is a unique one in the sense that survives these two criteria.

%%%%%%%%%%%%%%%%%%%%%%%%%%%%%%%%
%%%%%%%%%%%%%%%%%%%%%%%%%%%%%%%%
\section{More on effective metrics}
The most promising way of coupling the matter field is to couple it to only one metric, which guarantees the ghost freedom at the classical level \cite{Hassan:2011zd}. Furthermore, this property is not spoiled at the quantum level since the quantum corrections do not detune the special form of the potential interactions and contribute only in form of a cosmological constant \cite{deRham:2014naa}. If the matter field couples to both metrics simultaneously, the ghost degree of freedom reappears already at the classical level \cite{deRham:2014naa} and the quantum corrections detune the potential interactions at an unacceptable low scale.

The only way to couple the matter sector to the two metrics $g_{\mu\nu}$ and $f_{\mu\nu}$ simultaneously is through an effective composite metric, whose form is drastically restricted by the behaviour of quantum corrections. They have to be built in a way such that the quantum corrections do not detune the very specific potential structure in order not to reintroduce the ghost with an arbitrarily law scaling \cite{deRham:2014naa, Heisenberg:2014rka}. 

%%%%%%%%%%%%%%%%%%%%%%%%%%%%%%%%%%%%%%%%%%%
\subsubsection{Effective composite metrics in form of cosmological constants}
One way of guaranteeing that the quantum corrections coming from the matter loops do not detune the potential interaction is by imposing that they shall contribute additively in from of cosmological constants for $g_{\mu\nu}$ and $f_{\mu\nu}$, in other words the determinant of the effective metric is restricted to be
\begin{equation}\label{additionalDet}
\sqrt{-\det{g_{\rm eff}}}=\sqrt{-\det g}+\sqrt{-\det f}
\end{equation} 
This ensures that the quantum corrections do not destroy the naturalness property of massive gravity. We can solve this relation to find out that the form of the effective composite metric is forced to be \cite{Heisenberg:2014rka}
\begin{equation}\label{additiveGeffsol}
g^{\rm eff}_{\mu\nu}=\left( \sqrt{-\det g}+\sqrt{-\det f} \right)^{1/2} M_{\mu\nu}
\end{equation}
with the only requirement that $\det M=1$. This corresponds to a nine parametric solution. This is fulfilled by constructing $M$ as 
\begin{equation}\label{ansatzforM}
M_{\mu\nu}=\frac{1}{(-\det N)^{1/4}}N_{\mu\nu}
\end{equation}
where $N$ in principle can be arbitrary. The simplest would correspond to choosing $M_{\mu\nu}=\eta_{\mu\nu}$.

%%%%%%%%%%%%%%%%%%%%%%%%%%%%%%%%%%%%%%%%%%%%%%%
%%%%%%%%%%%%%%%%%%%%%%%%%%%%%%%%%%%%%%%%%%%%%%%
\subsubsection{Effective composite metrics in form of potential interactions}
The only other possible way of coupling matter fields to both metrics is through an effective composite metric which gives rise to quantum corrections in form of the allowed ghost-free potential interactions \cite{deRham:2014naa}. This would destroy the naturalness of the theory but at least would not reintroduce the ghostly degree of freedom. For this to happen, the determinant of the effective metric needs to fulfil this time the following requirement
\begin{equation}\label{potentialDetgeff}
\sqrt{-\det{\tilde{g}_{\rm eff}}}=\sqrt{-\det g}\det(\alpha \mathbbm 1+\beta X)
\end{equation}
where $X=\sqrt{g^{-1}f}$. The generic solution to this equation will dictate $g_{\rm eff}$ to be of the form \cite{Heisenberg:2014rka}
\begin{equation}\label{solgeffgen}
\tilde{g}^{\rm eff}_{\mu\nu}=  g_{\mu\rho}\left[ (\alpha \mathbbm 1+\beta X)^2\right]^\rho_\sigma \tilde{M}^\sigma_\nu
\end{equation}
with $\alpha$ and $\beta$ being arbitrary constants and with the restriction $\det \tilde{M}=1$. Again, the nine parametric solution is guaranteed with $\tilde{M}=\tilde{N}/\det(\tilde{N})^{1/4}$. The simplest case with $\tilde{M}= \mathbbm 1$ corresponds to
\begin{equation}\label{geffdRHR}
\tilde{g}^{\text{eff}(\tilde{M}= \mathbbm 1)}_{\mu\nu}=\alpha^2 g_{\mu\nu} +2\alpha\beta g_{\mu\rho}
\Bigl(\sqrt{g^{-1}f}\Bigr)^{\rho}_\nu +\beta^2 f_{\mu\nu}\,,
\end{equation}
which is a particularly interesting one. In \cite{deRham:2014naa} it was shown that the matter coupling with this effective metric preserve the ghost freedom up to the scale $\Lambda_3^3=m^2M_{\text{Pl}}$. 

 All these classes of effective composite metrics ensure that quantum contributions maintain the potential interactions and do not reintroduce the ghost with an arbitrarily law scaling. This at least guarantees the quantum stability under matter loops. Nevertheless, it is crucial to investigate whether they contain the Boulware-Deser ghost already at the classical level. Not all of the effective composite metrics will be ghost-free at the classical level and one has to study carefully which of them is promising to be ghost-free classically. This is the main goal of this work. We will investigate whether they reintroduce the ghost and whether they can be considered as an effective field theory and it turns out that there is only one unique effective metric, that fulfils our restrictions at the classical level.

%%%%%%%%%%%%%%%%%%%%%%%%%%%%%%%%
%%%%%%%%%%%%%%%%%%%%%%%%%%%%%%%%
\section{Mini-superspace }

First of all, the effective composite metrics introduced in the previous section need to pass the easiest case of mini-superspace.
If they are predestined to be non linear in the lapse already in the mini-superspace then we do not need to go through more
involved space-times. Therefore, let us assume the following Ansatz for the $g$ and $f$ metrics
\begin{eqnarray}
\ud s_g^2&=&g_{\mu\nu} \ud x^\mu \ud x^\nu = -n_g^2 \ud t^2 + a_g^2
\ud x^2\,, \nonumber\\ \ud s_f^2&=&f_{\mu\nu} \ud x^\mu \ud x^\nu =
-n_f^2 \ud t^2 + a_f^2 \ud x^2\,,
\end{eqnarray}
where $n_g$, $n_f$ and $a_g$, $a_f$ are the lapse functions and scale factors of the $g$ and $f$ metric respectively and they only depend on the cosmic
time $t$.  We will assume the same symmetries for $N$ in (\ref{ansatzforM}) and make a similar Ansatz 
\begin{equation}
\ud s_N^2=N_{\mu\nu} \ud x^\mu \ud x^\nu = -n_N^2 \ud t^2 + a_N^2
\ud x^2\,.
\end{equation}
Hence $M$ would be parametrised in the mini-superspace as 
\begin{equation}
M_{\mu\nu}=\left[-(n_N/a_N)^{3/2},(a_N/n_N)^{1/2}\delta_{ij} \right],
\end{equation}
with $\det M=1$ guaranteed. Without loss of generality let us consider a scalar matter field for concreteness that minimally couples to the effective metric
\begin{equation}
\mathcal{L}_\text{matter} =-\frac{1}{2}\sqrt{-g_{\text{eff}}}\left( g_{\text{eff}}^{\mu\nu}\partial_\mu\chi \partial_\nu\chi +2V(\chi) \right)
\end{equation}
with the effective metric (\ref{additiveGeffsol}) in the mini-superspace given by
\begin{equation}
g^{\text{eff}}_{\mu\nu}=(a_f^3n_f+a_g^3n_g)^{\frac12}\left[ -\left(\frac{n_N}{a_N}\right)^{\frac32} , \left(\frac{a_N}{n_N}\right)^{\frac12}
\delta_{ij} \right]\,.
\end{equation}
and its determinant $\sqrt{-g_{\text{eff}}}=a_f^3n_f+a_g^3n_g$. Our matter Lagrangian simplifies to
\begin{equation}
\mathcal{L}_\text{matter} =\frac12\sqrt{a_f^3n_f+a_g^3n_g}\dot{\chi}^2 \left(\frac{a_N}{n_N}\right)^{\frac32}-(a_f^3n_f+a_g^3n_g)V \,. \nonumber
\end{equation}
The conjugate momenta associated to the scalar field reads
\begin{align}
p_\chi = \sqrt{a_f^3n_f+a_g^3n_g}\dot{\chi} \left(\frac{a_N}{n_N}\right)^{\frac32} \,. 
\end{align}
After performing the Legendre transformation we obtain for the Hamiltonian:
\begin{equation}\label{hamiltonian1}
\mathcal{H}_{\text{matter}}=\frac12\frac{p_\chi^2}{\sqrt{a_f^3n_f+a_g^3n_g}} \left(\frac{n_N}{a_N}\right)^{\frac32}+(a_f^3n_f+a_g^3n_g)V  \,,
\end{equation}
We have to ensure that the Hamiltonian is linear in the lapses $n_g$ and $n_f$. The part coming from the potential term already fulfils this requirement. In order for the Hamiltonian to be linear in the lapses we have to impose conditions on $n_N$ in terms of $n_g$ and $n_f$ such that the term in front of the conjugate momenta in the Hamiltonian (\ref{hamiltonian1}) is linear in the lapses. Note that $n_N$ and $a_N$ are functions of the lapses and scale factors of the $g$ and $f$ metrics. The necessary condition is
\begin{equation}\label{condition1}
(a_f^3n_f+a_g^3n_g)^{-\frac12} \left(\frac{n_N}{a_N}\right)^{\frac32} = \text{linear in the lapses}
\end{equation}
At this stage, we see immediately that the simplest case with $M_{\mu\nu}=\eta_{\mu\nu}$ would not give rise to an allowed effective composite metric since the Hamiltonian would be highly non-linear in the lapses. Actually, upon closer investigation one finds out
that this equation (\ref{condition1}) has no real solutions for $n_N$ and $a_N$ in terms of the lapses and scale factors of the $g$ and $f$ metrics. This urges us to abandon the couplings constructed out of this class of effective metrics.

Let us as next investigate the mini-superspace of the second type of effective composite metrics in equation (\ref{potentialDetgeff}). Again, we parametrize $\tilde{N}$ such that $\tilde{M}^{\mu}_{\nu}=\left[(\tilde{n}_N/\tilde{a}_N)^{3/2},(\tilde{a}_N/\tilde{n}_N)^{1/2}\delta_{ij} \right]$. The effective metric (\ref{potentialDetgeff}) becomes this time
\begin{equation}
\tilde{g}^{\text{eff}}_{\mu\nu}=\left[ -(\alpha n_g+\beta n_f)^2\left(\frac{\tilde{n}_N}{\tilde{a}_N}\right)^{\frac32} , (\alpha a_g+\beta a_f)^2\left(\frac{\tilde{a}_N}{\tilde{n}_N}\right)^{\frac12}
\delta_{ij} \right] \nonumber
\end{equation}
and its determinant $\sqrt{-\tilde{g}_{\text{eff}}}=(\alpha n_g+\beta n_f) (\alpha a_g+\beta a_f)^3$. The matter Lagrangian this time reads
\begin{eqnarray}
\tilde{\mathcal{L}}_\text{matter}&=&\frac12 (\alpha a_g+\beta a_f)^3 \frac{\dot{\chi}^2}{(\alpha n_g+\beta n_f)} \left(\frac{\tilde{a}_N}{\tilde{n}_N}\right)^{\frac32} \nonumber\\
&&-(\alpha n_g+\beta n_f) (\alpha a_g+\beta a_f)^3V \,. \nonumber
\end{eqnarray}
Similarly, the conjugate momenta associated to the scalar takes now the form
\begin{align}
\tilde{p}_\chi = \frac{(\alpha a_g+\beta a_f)^3}{(\alpha n_g+\beta n_f)} \dot{\chi} \left(\frac{\tilde{a}_N}{\tilde{n}_N}\right)^{\frac32} \,, 
\end{align}
and the Hamiltonian becomes
\begin{eqnarray}\label{hamiltonian2}
\tilde{\mathcal{H}}_{\text{matter}}=\frac12 \frac{(\alpha n_g+\beta n_f)}{(\alpha a_g+\beta a_f)^3} \tilde{p}_\chi^2 \left(\frac{\tilde{n}_N}{\tilde{a}_N}\right)^{\frac32} \nonumber\\
+(\alpha n_g+\beta n_f) (\alpha a_g+\beta a_f)^3V  \,.
\end{eqnarray}
We see immediately that for the simplest case $\tilde{M}^\mu_\nu=\delta^\mu_\nu$ which corresponds to the $g_{\text{eff}}$ of equation (\ref{geffdRHR}) proposed in \cite{dRHRa} gives rise to a Hamiltonian linear in the lapses, which coincides with the findings in \cite{dRHRa} 
\begin{eqnarray}\label{hamiltoniandelta}
\tilde{\mathcal{H}}_{\text{matter}}^{\tilde{M}=\mathbbm 1}= \frac{\tilde{p}_\chi^2 (\alpha n_g+\beta n_f)}{2(\alpha a_g+\beta a_f)^3} 
+(\alpha n_g+\beta n_f) (\alpha a_g+\beta a_f)^3V \nonumber   \,.
\end{eqnarray}
The question is now whether we can construct $\tilde{n}_N$ different than 1, such that the Hamiltonian remains linear in the lapses. Since we already have the linear term $(\alpha n_g+\beta n_f)$ in front of $\tilde{n}_N^{3/2}$, the only way to guarantee the linearity is unfortunately only by imposing $\tilde{n}_N=1$. Among all these effective metrics, (\ref{geffdRHR}) with $\tilde{M}^\mu_\nu=\delta^\mu_\nu$ is the only one that survives our criteria in the mini-superspace and hence is very special.
%%%%%%%%%%%%%%%%%%%%%%%%%%%%%%%%%%%%%%%%%%%%%
%%%%%%%%%%%%%%%%%%%%%%%%%%%%%%%%%%%%%%%%%%%%%
\section{Decoupling limit}
We will dedicate this section to the decoupling limit analysis of the matter couplings to the two types of effective metrics and confirm that only for the unique effective metric with $\tilde{M}^\mu_\nu=\delta^\mu_\nu$ the BD ghost remains absent in the decoupling limit. This will strengthen our results of the mini-superspace. For this purpose we will focus on the helicity-0 mode of the massive graviton. As usual we will first restore broken diffeomorphism invariance via Stueckelbergalization of the $f$ metric
\begin{equation}
 \tilde f_{\mu\nu} = f_{ab}\partial_\mu \phi^a \partial_\nu \phi^b\,,
\end{equation}
in terms of the helicity-0 $\pi$ and -1 $A$ counterparts of the Stueckelberg fields,
\begin{equation}
\phi^a=x^a-\frac{A^a}{mM_g} -\frac{f^{ab}\partial_b \pi}{\Lambda^3_3}\,,
\end{equation}
with $\Lambda^3_3=M_g m^2$. The decoupling limit corresponds to the scaling where $M_{g}, M_{f} \to \infty$, while $\Lambda_3$ kept fixed. We will set the contributions coming from the helicity-1 mode to zero and concentrate only on the contributions of the helicity-0 mode $\pi$. We have to make sure that there are no higher derivative terms of the helicity-0 interactions after using the equation of motion of the matter field $\chi$. We have
\begin{eqnarray}
\label{eq:fDL}
g_{\mu\nu}&=&\eta_{\mu\nu} \nonumber\\
f_{\mu\nu} &=&  \left(\eta_{\mu\nu}-\Pi_{\mu\nu}\right)^2\,,
\end{eqnarray}
with the short-cut notation $\Pi_{\mu\nu} \equiv \partial_\mu \partial_\nu \pi/\Lambda^3$. The first type of effective metric (\ref{additiveGeffsol}) in the decoupling limit can be expressed as
\begin{eqnarray}
g_{\mu\nu}^{\text{eff}}=\mathcal{C} M_{\mu\nu}(\Pi)
\end{eqnarray}
where $\mathcal{C}=\left[1+\det(\eta_{\rho\sigma}- \Pi_{\rho\sigma}) \right]^{\frac12}$ and we assumed that $M$ can in principle depend in a non-trivial way on $\Pi$.
The equation of motion for the matter field $\chi$ 
\begin{equation}\label{stressenergytensorConserved}
\nabla_\mu^{g_{\text{eff}}}T^{\mu\nu}_{g_{\text{eff}}}=0 \to \partial_\mu \left( \sqrt{-g_{\text{eff}}}T^{\mu\nu}_{g_{\text{eff}}} \right)=- \sqrt{-g_{\text{eff}}}\Gamma^{(g_{\text{eff}})\nu}_{\mu\rho}T^{\mu\rho}_{g_{\text{eff}}}
\end{equation}
corresponds to the conservation of the stress energy tensor. As next we can compute the equation of motion with respect to the helicity-0 field
\begin{eqnarray}\label{EOMforpi1}
\frac{\delta \mathcal{L}_{mat}}{\delta \pi}&=& \frac{\partial_\mu \partial_\nu}{\Lambda^3}\left( \frac{\delta \mathcal{L}_{mat}}{\delta g^{\text{eff}}_{\rho\sigma}} \frac{\delta g^{\text{eff}}_{\rho\sigma}}{\delta \Pi_{\mu\nu}} \right). \nonumber\\
\end{eqnarray}
Note that we can perform the following replacement in (\ref{EOMforpi1})
\begin{equation}
 \frac{\delta \mathcal{L}_{mat}}{\delta g^{\text{eff}}_{\rho\sigma}}=\frac{1}{2} T_{g_{\text{eff}}}^{\rho\sigma}\sqrt{-g_{\text{eff}}} \,,
\end{equation}
and furthermore we have
\begin{eqnarray}
\frac{\delta g^{\text{eff}}_{\rho\sigma}}{\delta \Pi_{\mu\nu}}= M_{\rho\sigma}(\Pi)\frac{\delta\mathcal{C}}{\delta \Pi_{\mu\nu} } 
+\mathcal{C} \frac{\delta M_{\rho\sigma}(\Pi)}{\delta \Pi_{\mu\nu}}.
\end{eqnarray}
We can explicitly compute the variation of $\mathcal{C}$ with respect to $\Pi$
\begin{eqnarray}
2\mathcal{C}\frac{\delta\mathcal{C}}{\delta \Pi_{\mu\nu} } &=& \eta^{\mu\nu}\left(6e_0(\Pi)+3\cdot 2!e_1(\Pi)+3!e_3(\Pi) \right) \nonumber\\
&-&3\Pi^{\mu\nu} \left(2e_0(\Pi)+2e_1(\Pi)+2!e_2(\Pi) \right)  \nonumber\\
&+&3\Pi^{\mu\alpha}(-2(1+e_1(\Pi))\Pi^\nu_\alpha +2 \Pi_\alpha^\beta \Pi_\beta^\nu)
\end{eqnarray}
with $e_n$ indicating the symmetric polynomials of $\Pi$. After applying the two derivatives, we immediately see that there is no way of cancelling the higher derivative terms even after taking into account the equation of motion for the matter field, hence all these effective metrics in the additive form \ref{additiveGeffsol} will give rise to ghostly degree of freedom
in the decoupling limit. This is in agreement with the mini-superspace analysis. The problem does not come from the presence of $M_{\mu\nu}(\Pi)$, but namely from the additive structure of the effective metrics. Even in the trivial case for $M$, the equation of motion for the helicity-0 mode contains higher derivatives. Concerning the 
second type of effective metrics (\ref{solgeffgen}), they are more special in the sense that they are constructed in a multiplicative way
and the chances of cancelation are higher then in the first type of effective metrics where $\sqrt{-g}$ and $\sqrt{-f}$ were added. In the mini-superspace we immediately saw that there was no way of making the Hamiltonian linear in the lapses unless $n_N$ and $a_N$ was
chosen to be $1$. In the decoupling limit, this fact is confirmed in the equation of motion for the helicity-0 mode where the only way of avoiding higher order derivative terms is by choosing the matrix $M$ to be unity. The effective metric of second type in the decoupling limit takes the form
\begin{eqnarray}
\tilde{g}_{\mu\nu}^{\text{eff}}=\left[ (\alpha+\beta)\eta_{\mu\rho}-\beta \Pi_{\mu\rho} \right]^2\tilde{M}^\rho_\nu(\Pi) \,,
\end{eqnarray}
and similarly the equation of motion for the helicity-0 mode this time reads
\begin{eqnarray}
\frac{\delta \mathcal{L}_{mat}}{\delta \pi}&=&  \frac{\partial_\mu \partial_\nu}{\Lambda^3}\left( \frac{\delta \mathcal{L}_{mat}}{\delta \tilde{g}^{\text{eff}}_{\rho\sigma}} \frac{\delta \tilde{g}^{\text{eff}}_{\rho\sigma}}{\delta \Pi_{\mu\nu}} \right) \nonumber\\
&=& -\frac{\partial_\mu \partial_\nu}{\Lambda^3} \Big( \beta  \sqrt{-\tilde{g}_{\text{eff}}}T^{\mu\sigma}_{\tilde{g}_\text{eff}}((\alpha+\beta)\delta^\nu_\rho-\beta\Pi^\nu_\rho) \tilde{M}_{\sigma}^\rho\nonumber\\
&-&\left. \frac12 \sqrt{-\tilde{g}_{\text{eff}}}T^{\rho\sigma}_{\tilde{g}_\text{eff}}((\alpha+\beta)\eta_{\rho\kappa}-\beta\Pi_{\rho\kappa})^2\frac{\delta \tilde{M}^\kappa_\sigma}{\delta \Pi_{\mu\nu}} \right) \nonumber  \,.
\end{eqnarray}
One sees immediately that the second line will give rise to higher order derivative interactions once we apply the derivatives on it, which we can not remove by using the equation of motion for the matter field $\nabla_\mu^{g_{\text{eff}}}T^{\mu\nu}_{g_{\text{eff}}}=0$. Concerning the first line, this has only the chance of cancelation if we impose $M^\rho_\nu=\delta^\rho_\nu$, in which case the higher order derivate terms cancel upon the use of (\ref{stressenergytensorConserved}) and the fact that the Christoffel symbols in this case simply becomes 
\begin{equation}
\Gamma^{\rho\,\tilde{g}_{\text{eff}}}_{\mu\sigma}
=-\tilde{g}_{\text{eff}}^{\rho\kappa}\bigl[(\alpha+\beta)\delta^\nu_\kappa
  -\beta\Pi^\nu_\kappa\bigr]\partial_\nu \Pi_{\mu\sigma}\,.
\end{equation}
The decoupling limit analysis together with the mini-superspace analysis indicate that there is a unique effective composite metric that does not excite the ghost in the decoupling limit and provides linear Hamiltonian in the lapses in maximally symmetric space-times. This corresponds exactly to the effective metric proposed in \cite{deRham:2014naa}
\begin{equation}\label{solgeffgen}
\tilde{g}^{\rm eff}_{\mu\nu}=  g_{\mu\rho}\left[ (\alpha \mathbbm 1+\beta X)^2\right]^\rho_\sigma \delta^\sigma_\nu
\end{equation}
which makes it very unique and special.
%%%%%%%%%%%%%%%%%%%%%%
%%%%%%%%%%%%%%%%%%%%%
\section{Conclusions}
In this work we investigated the two types of effective composite metrics proposed in \cite{Heisenberg:2014rka}. They are special in the sense that they guarantee a nice behaviour at the quantum level. Coupling the matter fields to these type of effective metrics generate quantum corrections in form of cosmological constants for the two metrics or to renormalize the potential interaction without detuning their relative tuning. However, once we ensure this property at the quantum level, we have to impose the consistency at the classical level as well, which was the aim of this work. First of all, the mini-superspace analysis showed immediately that there is no way of providing a Hamiltonian linear in the lapses for the first type of effective metrics, which give rise to quantum corrections in form of cosmological constants. This statement rules out this entire class of effective metrics as a viable classical couplings. We also saw that the equation of motion for the helicity-0 mode in the decoupling limit is predestined to give rise to higher order derivatives that we can not cancel. Concerning the second type of composite metrics, the mini-superspace analysis showed that the Hamiltonian remains linear only if we force $\tilde{M}= \mathbbm 1$. The ghost freedom in the decoupling limit requires the same condition. The analysis of this work shows the uniqueness of the effective metric (\ref{solgeffgen}) proposed in \cite{deRham:2014naa}.\\
In the very latest stage of this work we became aware of similar conclusions made in \cite{deRham:2015cha} in the vielbein formulation and in \cite{Huang:2015yga} in the metric formulation. Our analysis is complementary to these works and is in complete agreement. 

\acknowledgments L.H. wishes to
acknowledge the African Institute for Mathematical Sciences in
Muizenberg, South Africa, for hospitality and support at the latest
stage of this work.

\bibliography{geffs_LB_LH.bib}

\end{document}